\documentclass[
    ,final            
  ]
  {aipproc}

\layoutstyle{6x9}

\begin{document}

\begin{flushright}
OU-HET-571, \,December, 2006
\end{flushright}
\vspace{-10mm}

\title{Generalized Form Factors, Generalized Parton Distributions
and Spin Contents of the Nucleon}


\author{M. Wakamatsu}{
  address={Department of Physics, Faculty of Science, 
  Osaka University, Osaka 560-0004, JAPAN}
}



\begin{abstract}

A model independent prediction is given for the nucleon spin
contents, based only upon a reasonable theoretical postulate
on the isosinglet anomalous gravitomagnetic moment of the nucleon.

\end{abstract}

\maketitle


\section{Introduction}

If the intrinsic quark spin carries little
of the total nucleon spin, what carries the rest of the nucleon
spin? That is the question we must answer.
Quark orbital angular momentum (OAM) $L^Q$? Gluon OAM $L^g$?
Or gluon polarization $\Delta g$?
Several remarks are in order here. 
The recent COMPASS measurement of the quasi-real photoproduction
of high-$p_T$ hadron pairs indicates that $\Delta g$ is likely to be
small \cite{COMPASS:2006G}.
There also appeared an interesting paper by Brodsky and
Gardner, in which, based on the conjecture on the relation between the
Sivers mechanism and the quark and gluon OAM, it was argued
that small single-spin asymmetry observed by the COMPASS collaboration
on the deuteron target is an indication of
small gluon OAM \cite{BG:2006}.

On the other hand, the importance of quark OAM was pointed out many
years ago based on the
chiral soliton picture of the nucleon : first within the
Skyrme model \cite{BEK:1988}, second within the Chiral Quark
Soliton Model \cite{WY:1991}.
According the latter, the dominance of
quark OAM is intimately connected with collective motion of quarks
in the rotating hedgehog mean field.
The CQSM predicts at the model energy scale around $600 \,\mbox{MeV}$
that $\Delta \Sigma$ is around 0.35, while $2 \,L_q$ is around 0.65.
The CQSM also well reproduces the spin structure functions for
the proton, neutron and the deuteron \cite{WK:1999},\cite{Waka:2003}.
Very recently, new measurement of the deuteron spin structure
function $g_1^d (x,Q^2)$ was reported by the COMPASS
group \cite{COMPASS:2005D}.
The precise measurement of $g_1^d (x,Q^2)$ is very important, since,
aside from a small effects of $s$-quark polarization as well as the
nuclear effects, it is just proportional to the iso-singlet spin
distribution, the integral of which gives the intrinsic quark-spin
contribution to the nucleon spin.
The left panel of Fig.1 show the comparison between our predictions
for $x \,g_1^d (x,Q^2)$ given several years ago and the new COMPASS
data as well as the old SMC data, while the right panel of Fig.1
gives a similar comparison
for the quantity $g_1^N (x,Q^2) = g_1^d (x,Q^2) / (1 - 1.5 \omega_D)$ 
with $\omega_D = 0.05 \pm 0.01$.
One can clearly see that the new COMPASS data became closer to our
prediction especially at the smaller $x$ region.
This means that the CQSM reproduces not only the magnitude of
$\Delta \Sigma$ but also its Feynman $x$-space distribution as well.

\begin{figure}
  \includegraphics[height=.3\textheight]{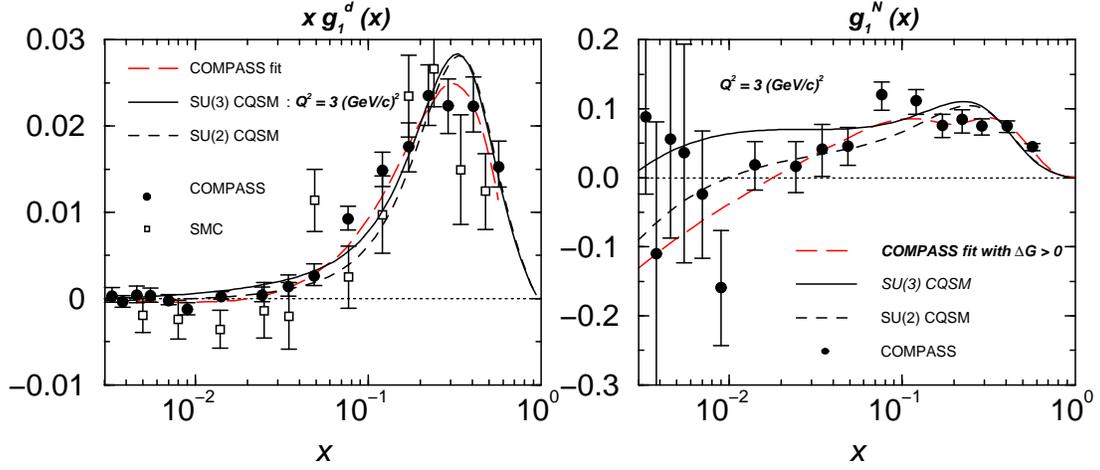}
  \caption{The predictions of the SU(2) and SU(3) CQSM in
  comparison with the new COMPASS data for $x \,g_1^d (x)$
  (left panel) and $g_1^N (x)$ (right panel). The old SMC
  data are also shown.}
\end{figure}

The recent noteworthy development is the
possibility of direct measurement of $J_q$ and $L_q$, through the
generalized parton distribution functions appearing in high-energy
deeply virtual Compton scatterings and deeply virtual meson productions.
What plays a crucial role here is the celebrated Ji's angular momentum
sum rule.

\section{Generalized form factors and Ji's angular momentum sum rule}

The generalized form factors $A_{20} (t)$ and $B_{20}(t)$ for the
quarks and gluons are defined as the nucleon nonforward matrix
elements of QCD energy momentum tensor $T^{\mu \nu}$.
The famous Ji's sum rule relates the total angular momentum carried
by quarks and gluons to the forward limits of these generalized
form factors : 
\begin{eqnarray*}
 J^{u + d} &=& \frac{1}{2} \,
 \left[ \,A_{20}^{u+d} (0) 
 \ + \ 
 B_{20}^{u+d}(0) \,\right], \ \ \ 
 J^g \ = \ \frac{1}{2} \,
 \left[ \,A_{20}^g (0) 
 \ \ + \ \ 
 B_{20}^g (0) \,\right] .
\end{eqnarray*}
Here, the first $A$ part reduces to the total momentum
fractions of quarks and gluons, 
$A_{20}^{u+d}(0) = {\langle x \rangle}^{u+d}, \,
A_{20}^g (0) = {\langle x \rangle}^g$,
while the second $B$ part represents the quark and gluon contribution
to the anomalous gravitomagnetic moment (AGM) of the nucleon.
An important observation is that the total nucleon AGM
identically vanishes : $B^{u+d}_{20} (0) + 
B^{g}_{20} (0) = 0$. This follows from the two sound
identities, i.e. the total momentum sum rule 
${\langle x \rangle}^{u+d} + {\langle x \rangle}^g = 1$ and 
the total angular momentum sum rule $J^{u+d} + J^g = 1/2$ 
of the nucleon. 
Here, there are two possibilities. In the 1st case,
the quark and gluon contribution to the nucleon AGM vanishes
separately. In the 2nd case, both have sizable magnitude with opposite
signs.
Fig.2 shows the theoretical predictions for the relevant
AGM form factors \cite{WN:2006}.
As one sees, both of LHPC lattice QCD simulation \cite{LHPC:2004}
and the CQSM predicts very small AGM consistent with zero.


\begin{figure}
  \includegraphics[height=.25\textheight]{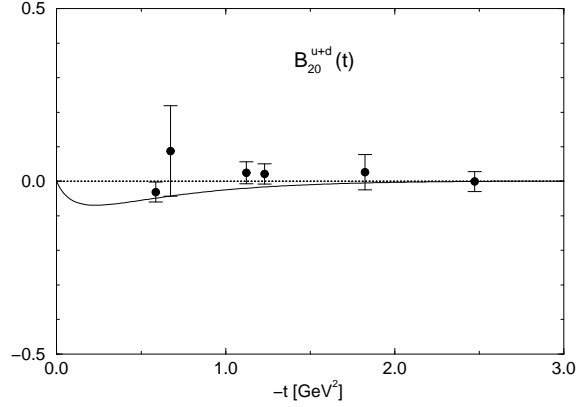}
  \caption{The CQSM prediction for $B_{20}^{u+d} (t)$ compared with
  LHPC lattice prediction.}
\end{figure}

The nucleon AGM is a measurable quantity, since it is given as
the Mellin moment of unpolarized GPD $E (x,\xi,t)$.
For model calculation,
the spin decomposition is most conveniently done in the Breit frame.
In this frame, the following combinations of GPDs $H$ and $E$ appear :
$H_E (x,\xi,t) = H(x,\xi,t) \, + \,
\frac{t}{4 \,M_N^2} \, E(x,\xi,t), \,
E_M (x,\xi,t) = H(x,\xi,t) \ + \ E(x,\xi,t)$,
which corresponds to the famous Sachs decomposition of
the nucleon electromagnetic form factors. 
Mellin moment sum rules relevant to our discussion are as follows.
The 1st moment of GPD $E_M$ reduces to the isoscalar combination
of magnetic moment, which consists of canonical part and anomalous
part. 
\begin{eqnarray*}
 \int_{-1}^1 \,E_M^{u+d}(x,0,0) \,dx
 &=& 
 \int_{-1}^1 \,H^{u+d}(x,0,0) \,dx \ + \ 
 \int_{-1}^1 \,E^{u+d}(x,0,0) \,dx \\
 &=& \  (\,N^u + N^d \,) 
 \ + \ (\,\kappa^u + \kappa^d \, ) \ = \ 
 \mu^u + \mu^d .
\end{eqnarray*}
On the other hand, the 2nd moment gives the total angular
momentum carried by quarks, which is also made up of canonical part
and anomalous part.
\begin{eqnarray*}
 \int_{-1}^1 \,x \,E_M^{u+d}(x,0,0) \,dx
 &=& 
 \int_{-1}^1 \,x \,H^{u+d}(x,0,0) \,dx \ + \ 
 \int_{-1}^1 \,x \,E^{u+d}(x,0,0) \,dx \\
 &=& \ 
 (\,\langle x \rangle^u + \langle x 
 \rangle^d \,) 
 \ + \ 
 B_{20}^{u+d} (0) \ = \ 
 2 \, ( \,J^u + J^d \, ) .
\end{eqnarray*}
An important fact to remember here is that the isoscalar anomalous
magnetic moment of the nucleon is very small, which means that
the $x$ integral of $E^{u+d} (x,0,0)$ is small. This also indicates
that $B^{u+d}_{20} (0)$, obtained as a $x$-weighted integral of
$E^{u+d} (x,0,0)$, would be even smaller, assuming a smooth behavior
of $E^{u+d} (x,0,0)$. Curiously, the smoothness
assumption of $E^{u+d} (x,0,0)$ is not respected by the CQSM, due to
the effects of polarized Dirac sea. Nevertheless, the above
conjecture, i.e. the smallness or the absence of the net
quark contribution to the nucleon anomalous AGM is realized in a
nontrivial manner. (See ref.\cite{WN:2006}, for more detail.)

In view of the observation above, let us assume smallness of
$B_{20}^Q (0)$ and $B_{20}^g (0)$,
and set them to 0, for simplicity. We are then led to surprisingly
simple relations, i.e. the proportionality of the
the linear and angular momentum fractions for both of quarks
and gluons :
\begin{eqnarray*}
 J^Q \ = \ \frac{1}{2} \,\,{\langle x \rangle}^Q,
 \ \ \  
 J^g \ = \ \frac{1}{2} \,\,{\langle x \rangle}^g .
\end{eqnarray*}
Important facts to remember here are that the quark and gluon momentum
fractions are already well determined through the empirical
PDF fits. Furthermore, this proportionality relations holds
scale-independently, since $J^Q$ and $\langle x \rangle^Q$ and also
$J^g$ and $\langle x \rangle^g$ obey exactly the same evolution
equations.
There also exist phenomenological fits for the longitudinally
polarized PDFs, which contains the information for $\Delta \Sigma$
and $\Delta g$, although with larger uncertainties, compared with
the unpolarized case.
These information are enough to determine the quark and gluon OAM
as functions of energy scale.
Imagine now that the answers obtained in this way at
$Q^2 = 4 \,\mbox{GeV}^2$ are evolved down to low energy model scale.
Around the energy scale of $600 \,\mbox{MeV}$, one would then obtain
the following estimate for the nucleon spin contents :
\begin{eqnarray*}
 2 \,J^{u+d} \ = \ {\langle x \rangle}^{u+d} 
 &\simeq& 
 (0.75 \sim 0.8), \ \ \
 2 \,J^g \ = \ \langle x \rangle^g \ \simeq \ 
 (0.2 \sim 0.25), \\
 \Delta \Sigma &\simeq& (0.2 \sim 0.3), \ \ \ \ \ 
 2 \,L^{u+d} \ \simeq \ 
 (0.45 \sim 0.6) .
\end{eqnarray*}
which indicates that nearly half of the
nucleon spin comes from the quark OAM.


%

\section{Summary and Conclusion}

$L_Q$ or $\Delta g$ ? There has been long-lasting dispute over this
issue. Here, we have shown that, based only upon a reasonable
theoretical postulate, i,e. the smallness of the net quark
contribution to the nucleon AGM, we are
naturally led to the following conclusion for the nucleon spin
contents. At the low energy scale around $600 \,\mbox{MeV}$,
about $30 \%$ of the nucleon spin is due to the intrinsic quark
spin, whereas about $(40 \sim 50) \,\%$ comes from the quark OAM.
The remaining $(20 \sim 30) \,\%$ of the nucleon spin is likely to
be carried by the gluon fields. The last statement is not inconsistent
with the recent observation by Brodsky and Gardener, since what would
be related to the Sivers mechanism is the anomalous part of the gluon
OAM. On the other hand, our postulated identity
$2 \,J^g \simeq {\langle x \rangle}^g$ implies that this gluon
angular momentum comes totally from its canonical orbital motion,
not from the anomalous contribution related to the GPD $E(x,\xi,t)$.

%



\bibliographystyle{aipproc}   


%
%
%
%

\end{document}